\documentstyle[twocolumn,prl,aps,epsfig]{revtex}

\begin{document}
\title{
Interaction induced collapse of a section of the Fermi sea in  
	     in the zig-zag Hubbard ladder
       }

\author{Kay Hamacher$^{(1,3)}$, Claudius Gros$^{(2)}$
        and Wolfgang Wenzel$^{(3)}$
       }
\address{$^{(1)}$ Institut f\"ur Physik, Universit\"at Dortmund,
           44221 Dortmund, Germany.}

\address{$^{(2)}$Fakult\"at 7, Theoretische Physik,
 University of the Saarland,
66041 Saarbr\"ucken, Germany.}

\address{$^{(3)}$
Forschungszentrum Karlsruhe,
Institut f\"ur Nanotechnologie,
Hermann-von-Helmholtz-Platz 1,
76344 Eggenstein-Leopoldshafen, Germany.}

\maketitle

\begin{abstract}
Using the next-nearest neighbor (zig-zag) Hubbard chain as an one
dimemensional model, we investigate the influence of interactions on
the position of the Fermi wavevectors with the density-matrix
renormalization-group technique (DMRG). For suitable choices of the
hopping parameters we observe that electron-electron correlations
induce very different renormalizations for the two different Fermi
wavevectors, which ultimately lead to a complete destruction of one
section of the Fermi sea in a quantum critical point.
\end{abstract}
\maketitle


{\bf Introduction} -
Many aspects of the low-energy physics of an electronic system are
influenced by the shape of its Fermi surface and the occupation of
nearby states. Associated observables may vary strongly with the
temperature and the strength of the interaction of the
electrons. Changes in the Fermi-surface geometry induce magnetic and
other instabilities. Electron-electron interactions induce a momentum
dependent softening of the Fermi surface, which is responsible for a
wide variety of phenomena, such as the loss of magnetic order or
high-temperature superconductivity. In systems where the
Fermi surface is not simply connected, interactions may lead to a
partial or total collapse of parts of the Fermi surface at a quantum
critical point and an associated drastic change in the physical
properties of the system. It is therefore important to study the
renormalization of the individual Fermi-surface sections under the
influence of electron-electron correlations in competition with
frustrating interactions.

Recent investigations suggested a spontaneous, interaction induced
deformation of the Fermi surface of the of the 2D t-J
model~\cite{him00} as well as of the (extended) Hubbard
model~\cite{gro94,yod99,val00,mai01}. Because the effect of strong
electronic interactions is difficult to study in two- or
three-dimensional systems, these studies are confined either
to the weak coupling limit or involve extensive numerical
studies of effective lattice models. Many
aspects regarding the Fermi surface
renormalization for strong interactions remain presently
unclear.

The development of the density matrix renormalization group now offers
a reliable --- though numerically involved --- approach to rigorously
study the effect of strong interaction in one dimension. In this
letter we report a study of the renormalization of the Fermi
wavevector, or Fermi point, in the Hubbard chain with next-nearest-neighbor
interactions as a one-dimensional model for Fermi-surface
renormalization. We confirm results of a recent renormalization group
(RG) analysis for the Fermi point renormalization in weak
coupling. Fermi points close to a saddle-point of the momentum
distribution function $n(k)$ are predicted to shift towards the
van-Hove singularity. A similar renormalization has been suggested for
two-dimensional systems~\cite{him00,val00}. For strong interactions,
the pocket of the Fermi-sea near the saddle-point may be emptied
completely, a hypothesis
\cite{lou01} which can not be verified in weak-coupling. In order to
study this prediction rigorously, we have performed a DMRG-study of
the $t_1-t_2$ Hubbard model for rings with periodic boundary
conditions, evaluating the momentum distribution function $n(k)$
directly for rings with up to 80 sites. For larger interaction
strength we find a novel interaction-induced quantum critical point
that is associated with the collapse of a Fermi sea pocket.
This phenomenon is analogous to the opening of a pseudogap,
found experimentally in the high-T$_C$-cuprates~\cite{kam00},
near the saddlepoint of the electronic dispersion.

\begin{figure}[t]
\centerline{\epsfig{file=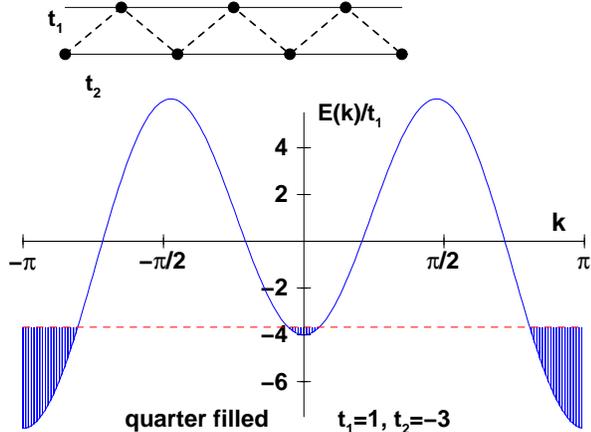,width=0.9\columnwidth}}
\smallskip
\caption{\label{dispersion}
Top: Illustration of the $t_1$-$t_2$ zig-zag ladder,
the dashed/full lines correspond to $t_1/t_2$ bonds.
Bottom: Illustration of the $U=0$ dispersion relation
$E(k)=2\,t_1\cos(k)+2\,t_2\cos(2k)$ for the
case $t_1=1$, $t_2=-3$. Also show (horizontal dashed line)
is the Fermi-energy  $E_f=-3.6736\,t_1$ for the
quarter-filled case.
}
\end{figure}

The Hamiltonian of the zig-zag Hubbard ladder (see Fig.\
(\ref{dispersion}), top) is given as:
\begin{eqnarray}
H & =&
\sum_{n,\sigma\atop\Delta n=1,2}
t_{\Delta n}^{\phantom{\dagger}}
\left(
c_{n+\Delta n,\sigma}^\dagger c_{n,\sigma}^{\phantom{\dagger}}
+ {\rm h.c.}\right)  \nonumber \\
&&\quad\qquad \,+\, U\sum_{n}
c_{n,\uparrow}^\dagger c_{n,\uparrow}^{\phantom{\dagger}}
c_{n,\downarrow}^\dagger c_{n,\downarrow}^{\phantom{\dagger}}~,
\label{def_H}
\end{eqnarray}
where the $c_{n,\sigma}^\dagger$ ($c_{n,\sigma}^{\phantom{\dagger}}$)
are Fermion-creation (destruction) operators on site $n$ and spin
$\sigma=\uparrow,\downarrow$. For appropriate choices of the
parameters this model describes either the low-energy properties of
Hubbard-ladders~\cite{lou01} or half-filled edge-sharing double-chain
materials like SrCuO$_2$~\cite{ric93} or LiV$_2$O$_5$~\cite{val01} for
which the next-nearest neighbor hopping $t_2$ is expected to be
substantial.

Generically, the variation of the two hopping parameters ($t_{1,2}$)
allows changes in the position of the Fermi point in the non-interacting
limit. When the hopping $t_2$ is large enough, the Fermi-sea will
split into two separate parts, see Fig.\ (\ref{dispersion}). According
to a recent RG analysis for the case of two separate
Fermi-seas\cite{lou01} the Fermi points $k_F^{(i)}$ ($i=1,2$) are
renormalized as:
\begin{equation}
-\Delta k_F^{(1)} = - \Delta
k_F^{(2)} = \Delta k\ \simeq\
{U^2\over2\pi^2}\,
{v_2-v_1\over(v_1+v_2)^2}\,{\Lambda_0\over v_1 v_2}~.
\label{Dk_RG}
\end{equation}
In weak coupling, the Fermi point-shift $\Delta k$ depends only on the
respective Fermi-velocities $v_{i}$ and on the initial momentum
cut-off $\Lambda_0$.  Eq.\ (\ref{Dk_RG}) is rigorous in the weak
coupling limit.


{\bf DMRG} - The evaluation of $n(k)$ as the Fourier transform of the
correlation function:
\begin{equation}
n_\sigma(k)\ =\ {2\over L}\sum_{n,n'=1}^L
\cos(k(n-n'))\, \langle\,
c_{n,\sigma}^\dagger
c_{n',\sigma}^{\phantom{\dagger}}\, \rangle
\label{nk_real}
\end{equation}
is numerically difficult and costly in the framework of the
DMRG~\cite{dmrgpaper,dmrgbuch}, in particular for periodic boundary
conditions~\cite{qin95}.
For large interaction strength, the suppression of double
occupancy reduces the size of the relevant Hilbert space and improves
the convergence properties of the DMRG procedure~\cite{diss}. We have
therefore decided to study the case of quarter-filling, where the
renormalization of the Fermi point occurs at relatively large values
of the interaction $U$. We chose $t_2/t_1=-3$ as a good set of
parameters to observe the Fermi point renormalization effects at
quarter-filling, since one of the Fermi-seas is small for this case
(see Fig.~(\ref{dispersion})) and since the weak-coupling RG predicts
gapless Luttinger-liquid behavior for these parameters~\cite{lou01}.

\begin{figure}[t]
\centerline{
\epsfig{file=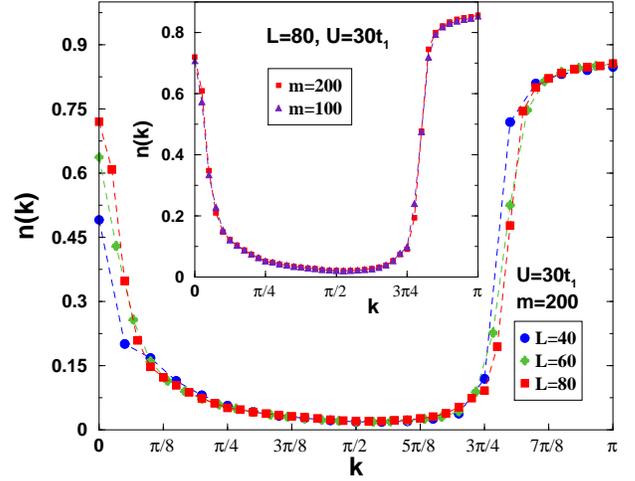,width=0.45\textwidth,angle=0}
}
\smallskip
\caption{\label{abb::conv}
Convergence of the correlation function $n(k)$
compared for different chain lengths $L$
and DMRG states $m$ (inset shows $L=80$) for a quarter-filled
system with $t_1=1$, $t_2=-3$. The lines are guides to the
eye.
}
\end{figure}

We checked for both the convergence of the variational energy and of
the correlation functions by increasing the number of 
DMRG states $m$ and found reasonable choices 
for the desired accuracy. The weight of the discarded fraction 
of the density matrix
varies between $1\times 10^{-4}$ to $2\times 10^{-5}$ depending on the
choice of system size and the number of states.
Fig.\ (\ref{abb::conv}) demonstrates the convergence of $n(k)$ both as a
function of lattice size and of the number of states $m$ in the DMRG
procedure. Additionally, we performed calculations for open chains and
verified that the behaviors of the real-space correlation function is
consistent with those for periodic systems.

{\bf Results} - Depending on the relative strength of the hopping
parameters $t_2/t_1$ and interaction strength $U/t_1$ the ground state
phase diagram of the $t_1-t_2$-model exhibits various instabilities towards
the formation of states with spin- and/or charge-gap~\cite{bal96,dau98}. We
have therefore investigated the possibility of instabilities towards
ferromagnetism and phase-separation by evaluating $\Delta E_F =
E(N_\uparrow+1,N_\downarrow-1)- E(N_\uparrow,N_\downarrow)$, where
$E(N_\uparrow,N_\downarrow)$ is the ground-state energy of the
quarter-filled system, and $\Delta E_{ph} =
E(N_\uparrow+1,N_\downarrow+1) + E(N_\uparrow-1,N_\downarrow-1)-
2E(N_\uparrow,N_\downarrow)$.  For the parameters considered here
no tendency towards an instability was found for any interaction
strength, in accordance with Ref.\cite{dau98}.

\begin{figure}[!t]
\centerline{
\epsfig{file=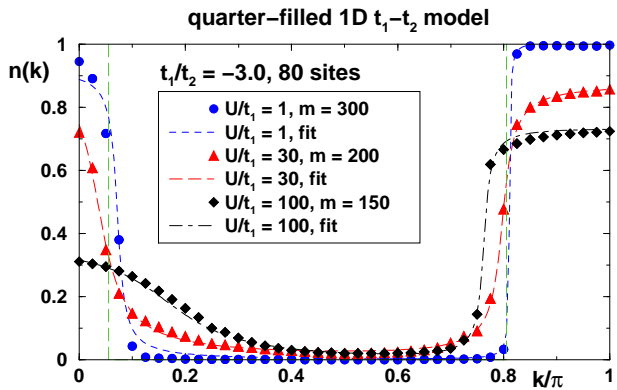,width=0.45\textwidth} 
}
\smallskip
\caption{\label{nk}
The DMRG-results (symbols) for the momentum distribution function for
a quarter-filled 80-site system with periodic boundary conditions. The
parameters are $t_1=1$, $t_2=-3$, and $U=1$ and $U=30$
respectively. The lines through the DMRG-data are fits by Eq.\
(\protect\ref{fit_nk}). The dashed vertical lines indicate the $U=0$
positions of the two Fermi-points.  }
\end{figure}

In Fig.\ (\ref{nk}) we present results for the momentum distribution
function for $U/t_1=1,30,100$. For $t_2=-3t_1$, quarter filling and
$U=0$ the two Fermi wave-vectors are $k_F^{(1)}= 0.055 \pi$ and
$k_F^{(2)} = 0.805 \pi$. The parameter $\alpha =
(v_F^{(1)}+v_F^{(2)})/(2 v_F^{(1)}) = 5.53$ entering the weak-coupling
RG-equations~\cite{lou01} indicates a gapless Luttinger-liquid phase
at small couplings.

For small U we clearly observe the existence of two Fermi-points
$k_F^{(1)}$ and $k_F^{(2)}$ (see Fig. 3), which are substantially
renormalized with growing interaction strength.

It is evident from the data presented in Fig.\ (\ref{nk}), that the
magnitude of the momentum distribution function inside the smaller
Fermi-sea is reduced substantially stronger by the interaction
than the one inside the large Fermi-point. This is a consequence of
the very different values for the respective Fermi-velocities, $v_1 =
0.658\,t_1$ and $v_2 = 6.620\,t_1$, which differ by one order of
magnitude. The energies of particle-hole excitations are $\sim v_F
\Delta k$, many more particle-hole excitations are therefore created
for the Fermi-sea pocket, which has the smaller $v_F=v_1$, resulting
in a substantial reduction in the occupation numbers $n(k)$ for
$k<k_F^{(1)}$.

In order to quantify the results of Fig. 3 we analyze the Fermi point
position as the point where the slope of the momentum-distribution
function is maximal. We find that our finite-lattice data for the
momentum distribution function $n(k)$ is approximated well by two
washed-out step-functions:
\begin{equation}
n(k)\ =\ a_0 + a_1\,{\rm atan}{k-k_F^{(1)}\over p_1}
             + a_2\,{\rm atan}{k-k_F^{(2)}\over p_2}~,
\label{fit_nk}
\end{equation}
The quality of this fit is illustrated by the lines in Fig.\
(\ref{nk}).

\begin{figure}[t]
\bigskip
\bigskip
\centerline{
\epsfig{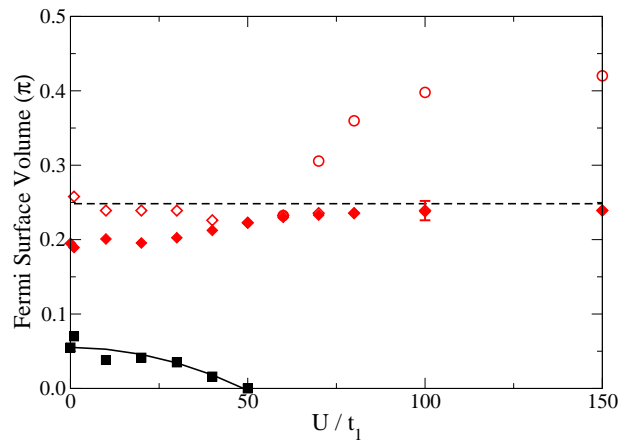}}
\smallskip
\caption{\label{dk}
The DMRG-results for the volume of the small ($k_F^{(1)}$, filled
squares, the line is a guide to the eye) and large ($\pi-k_F^{(2)}$,
filled diamonds) Fermi-sea and the total (open diamonds) volume of the
Fermi-sea (FSV). The latter is identical with the volume of the
remaining Fermi sea for $U > 50$. The dashed line indicates the
analytical value of the total FSV at $U=0$ (quarter filling). The open
circles indicate the total FSV under the assumption that a a small
Fermi point persists for $U>50$.  }
\end{figure}

Fig.\ (\ref{dk}) shows the dependence of $k_F^{(1/2)}$ as well as the
total volume of the Fermi sea as a function of $U$ (in units of
$\pi$).  The data clearly indicate the presence of a
quantum-critical-point with $U_c\approx 50t_1$, for which the smaller
Fermi sea collapses. The fit-parameter $k_F^{(1)}$ entering
(\ref{fit_nk}) denotes a locus of maximal slope in $n(k)$, which
corresponds to the Fermi-point position for $U<U_c$.  As the
calculations were done for fixed particle density, we expect the
Fermi-sea volume to be independent of the interaction-strength $U$
(Luttinger's theorem~\cite{yam97}). From Fig.\ (\ref{dk}) we see that
this expectation is approximatively fulfilled if we take
$(\pi-k_F^{(2)})+k_F^{(1)}$ for the volume of the Fermi-sea for $U <
U_c$ (open diamonds in Fig.\ (\ref{dk})) and and $\pi-k_F^{(2)}$ for
$U> U_c$. Some numerical uncertainty is observed, as usual in
numerical simulations, near the critical point.  The error bar in the
figure denotes the lattice induced finite-size error on the Fermi
point position. Within this error all data is compatible with the
exact result in the noninteracting limit.

As the data for $U = 100$ in Fig. 3 indicates, we have also maxima in
the slope of $n(k)$ corresponding to the remainder of the small
Fermi-sea for $U > U_c$. To distinguish between incoherent excitations
and a sharp Fermi point we have plotted the hypothetical total Fermi
sea volume (open circles in Fig. 4) under the assumption that these
inflection points also correspond to Fermi points.  As the figure
indicates this would lead to a massive violation of the Luttinger
theorem for all $U > U_c$. We thus conclude that $n(k)$ for small $k$
and $U > U_c$ can be consistently associated with contributions from
incoherent excitations to the momentum distribution function.


{\bf Conclusions} - In a one-dimensional model we have analyzed the
stability of Fermi-sea pockets with respect to electron-electron
interactions and found that sufficiently strong repulsive interactions
lead to a novel quantum critical point at which the one Fermi-sea
pocket is destroyed altogether.  We studied this quantum-phase
transition within the $t_1-t_2$ zig-zag ladder as a prototypical
model. We estimate the value of $U_c/t_1\approx 50$ for $t_2=3 t_1$ in
a quarter-filled system. We note that the nature and existence of this
transition does not depend on the specific choice of $t_1, t_2$ and
the filling, but that a manifold of transitions exist for appropriate
choices of these parameters. The conditions of the existence of the
transition are the existence of two different pockets of the Fermi sea
with a significantly higher Fermi velocity at the Fermi point of the
smaller pocket. In our study we have chosen the values of the
parameters such that the transition becomes amenable to quantitative
treatment with the DMRG, which converges progressively worse near
half-filled systems. This lead to a critical $U_c$ that seems to be
too large to be relevant for experimental realizations of
(\ref{def_H}), but we would like to point out that at least two
effects are likely to dramatically reduce the critical $U_c$:

(i) At half filling the $t_2-t_1$ model undergoes a Mott-Hubbard
transition for $t_2<t_1/2$. For $t_2\to t_1/2$ the critical
$U_c\to0$. It has been estimated \cite{lou01} that the RG-estimate for
the critical $U_c$ obtained from Eq.\ (\ref{Dk_RG}) is in agreement
with numerical results \cite{dau00,aeb01}.  We therefore expect the
critical $U_c$ for the destruction of the small Fermi-surface pocket
to reduce drastically near half-filling. Here we did not investigate
this region due to numerical difficulties.

(ii) In realistic systems the Coulomb-interaction will be
longer-ranged than the onsite-form assumed in the present study. It is
known, that longer-range contributions to the interaction increase
substantially the effect of the interaction on the Luttinger-liquid
parameters in the one-dimensional electron gas \cite{cre01,zhu01}.  We
expect a similar enhanced influence on the renormalization position of
the Fermi points.

We believe our rigorous findings for the one-dimensional case to be
relevant for two-dimensions. We note that the Fermi-surface
instability observed in the present study is qualitatively different
from the Pomeranchuk instability observed in
weak-coupling~\cite{hal00} for the 2D Hubbard model, since the
Pomeranchuk instability corresponds to a spontaneous breaking of the
tetragonal (C$_4$) symmetry due to enhanced scattering from one
saddle-point to another.  The Fermi-surface instability observed in
the present study is driven, on the other hand, by scattering between
`normal' and flat parts of the Fermi-surface.


{\bf Acknowledgments} -
KH gratefully acknowledges financial support by the Fonds der
chemischen Industrie, the BMBF and the Studienstiftung des dt.~Volkes.











\begin{thebibliography}{99}



\bibitem{him00} A.~Himeda and M.~Ogata,
                Phys. Rev. Lett. {\bf 85}, 4345 (2000)

\bibitem{gro94} C.~Gros and R.~Valent\'\i,
		Annalen der Phys. {\bf 3}, 460 (1994).

\bibitem{yod99} S.~Yoda and K.~Yamada,
                Phys. Rev. B. {\bf 60}, 7886  (1999)

\bibitem{val00} B.~Valenzuela and M.A.H.~Vozmediano,
                Phys. Rev. B. {\bf 63}, 153 103 (2001)

\bibitem{mai01} Th.A.~Maier, Th.~Pruschke and M.~Jarrell,
                Sissa, cond-mat/0111368;

\bibitem{lou01} K.~Louis, J.V.~Alvarez and C.~Gros,
                Phys. Rev. B {\bf 64 }, 11 3106 (2001).

\bibitem{kam00} See A.~Kaminski {\it et al.},
                Phys. Rev. Lett. {\bf 84}, 1788 (2000)
		and references therein.

\bibitem{ric93} T.M.~Rice, S.~Gopalan and M.~Sigrist,
                Europhys. Lett. {\bf 23}, 445 (1993).

\bibitem{val01} R.~Valent\'\i, T.~Saha-Dasgupta, J.V.~Alvarez,
            K.~Pozgajcic and C.~Gros,
            Phys. Rev. Lett {\bf 86}, 5381 (2001).

\bibitem{dmrgpaper}  S.R.~White, Phys. Rev. Lett. {\bf 69}, 2863 (1992)
                     and
                     S.R.~White, Phys. Rev. B {\bf 48}, 10 345 (1993).

\bibitem{dmrgbuch}  {{\em ``Density-Matrix Renormalization -- A New
                    Numerical Method in Physics''}},
                   {I.~Peschel, X.~Wang,
                   M.~Kaulke and K.~Hallberg (eds.)},
                  {Springer}, Lecture Notes in Physics,
                      {Berlin},  (1999).

\bibitem{qin95} S.~Qin, S.~Liang, S.~Su and L.~Yu,
                Phys. Rev. B {\bf 52}, R5475 (1995).

\bibitem{diss} K.~Hamacher, PhD thesis, Dortmund University (2001).

\bibitem{bal96} L.~Balents and M.P.A.~Fisher,
                Phys. Rev. B {\bf 53}, 12 133 (1996).

\bibitem{dau98} S.~Daul and R.M.~Noack,
                Phys. Rev. B {\bf 58}, 2635 (1998).

\bibitem{yam97} See M.~Yamanaka, M.~Oshikawa and I.~Affleck
	        Phys. Rev. Lett. 79, 1110 (1997) for a proof
		of Luttinger's theorem for one-dimensional systems.

\bibitem{dau00} S.~Daul and R.M.~Noack,
                Phys. Rev. B {\bf 61}, 1646 (2000).

\bibitem{aeb01} C.~Aebischer, D.~Baeriswyl and R.M.~Noack,
                Phys. Rev. Lett. {\bf 86}, 468 (2001).

\bibitem{cre01} C.E~Crefield, W.~H\"ausler and A.H.~MacDonald,
                Europhys. Lett. {\bf 53}, 221 (2001).

\bibitem{zhu01} A.K.~Zhuravlev and M.I.~Katsnelson
                Phys. Rev. B {\bf 64}, 033102 (2001)

\bibitem{hal00} C.J.~Halboth, W.~Metzner,
                Phys. Rev. Lett. {\bf 85}, 5162 (2000)

\end{thebibliography}
\end{document}